\documentstyle[aaspp4,12pt, psfig]{article}

\newcommand{\be}{\begin{equation}}
\newcommand{\ee}{\end{equation}}
\newcommand{\nn}{\mbox{} \nonumber \\ \mbox{} }
\newcommand{\ba}{\begin{eqnarray}}
\newcommand{\ea}{\end{eqnarray}}
\newcommand{\om}{\omega}
\newcommand{\Alfven}{Alfv\'{e}n }
\def\lo{\mathrel{\raise.3ex\hbox{$<$}\mkern-14mu\lower0.6ex\hbox{$\sim$}}}
\def\go{\mathrel{\raise.3ex\hbox{$>$}\mkern-14mu\lower0.6ex\hbox{$\sim$}}}

\begin{document}

\title{Rates of reconnection in pulsar winds}

\author{Maxim Lyutikov  $^{1,2,3}$}

\affil{$^1$ Physics Department, McGill University, 3600 rue University
Montreal, QC,  Canada H3A 2T8, \\
$^2$ Massachusetts Institute of Technology,
77 Massachusetts Avenue, Cambridge, MA 02139, \\
$^3$ CITA National Fellow}

\date{Received   / Accepted  }

\begin{abstract}
Using the  formulation of  relativistic reconnection by 
 Lyutikov \& Uzdensky (2002) we estimate the  upper  possible
 rates of reconnection
in pulsar winds using  Bohm prescription 
for  resistivity. We find
that the velocity of  plasma inflow into the  reconnection layer
may be   relativistic,  of the order  of the speed of light in the plasma
rest-frame.
This in principle
 may allow   efficient dissipation of the magnetic field energy
in the wind and/or   destruction
of the toroidal  magnetic  flux.
 The efficiency of reconnection
realized 
in pulsar winds remains an open question: it
 should  depend both on the microphysical properties of plasma
and on the  three-dimensional structure of the reconnection flow.
\keywords{stars: pulsars - plasmas - magnetic fields } 
\end{abstract}

\section{Introduction}

Magnetic reconnection is  a very important phenomenon
in many laboratory and astrophysical plasmas (Biskamp 2000,
Priest \& Forbes 2000). Most reconnection models are based on two
classical schemes suggested by  Sweet \& Parker  and  Petschek.
Recognition that 
magnetic reconnection
processes are also of great importance in high energy astrophysics,
 where  dynamic behavior
is often dominated by  super-strong magnetic fields,
with  energy density $B^2/(8 \pi)$ larger than the rest energy of the matter
$\rho c^2$,
has led Lyutikov \& Uzdensky (2002) to formulate the
relativistic Sweet-Parker reconnection.
In this Letter we apply the model of  relativistic reconnection to the
problem of energy conversion in pulsar winds.

The $\sigma$ paradox ($\sigma$ is conventionally defined as the ratio
 of Poynting to particle fluxes or, equivalently, as
a ratio of the magnetic  to plasma energy density) 
is a  long-standing problem in pulsar physics
(Rees \& Gunn 1974, Kennel \& Coroniti 1984). Models of the pulsar
magnetosphere (Goldreich \& Julian 1969,
 Arons \& Scharlemann 1979, Ruderman \& Sutherland 1975)
 predict that  near the light cylinder  most of the spin-down luminosity of
a pulsar
 should be in a form of Poynting flux, $\sigma \gg 1$.
On the other hand,  modeling of the dynamics of the Crab  nebula
(and other PWNs like W44, Shelton et al. 1999, 
 Vela,  Pavlov et al. 2001, PWN around PSR B1509-58, Gaensler et al. 2002)
gives a low value of $\sigma$ at what is commonly believed to be
a  reverse shock - strongly
magnetized  flows cannot  match the boundary conditions
 (Rees \& Gunn 1974, Kennel \& Coroniti 1984).

Most promising resolution of the $\sigma$ paradox invokes
 internal
dissipation of the magnetic fields in the equatorial flow
due to break down of MHD approximation
 (Coroniti 1990, Michel 1994,  Melatos \& Melrose 1996).
In the equatorial plane of an oblique rotator,
 the MHD wind forms  striped structures, in which 
  alternating
$B_\phi$ regions  are separated  by
 current sheets.
As  plasma flows out from the pulsar, the plasma density
decreases in proportion to $r^{-2}$ reaching a critical radius, where
it becomes less than the  critical charge density $n_{\rm crit} \sim r^{-1}$
 required to  carry
 the current. 
 Beyond this  
limiting radius  MHD approximation breaks down. 
Coroniti (1990) has argued that 
such
a breakdown of MHD would lead to  effective dissipation of the
field. This process was called reconnection since it is  supposed to 
 destroy  magnetic field and transfer  energy to particles, 
similar to the effects of reconnection
 in a classical (e.g. Solar physics) sense.
Later Michel (1994)  extended
the Coroniti's model, arguing that if  reconnection is efficient,
the structure of the wind nebula may be very different  from the
 ideal MHD one
envisioned by the Kennel \& Coroniti (1984). In particular,
the flow may not have a reverse shock, but 
 may simply decelerate smoothly to match nebular boundary conditions.
This model has been criticized by  Lyubarsky \& Kirk (2001) (LK afterward)
who argued that  acceleration of the flow, resulting from 
extra pressure released during reconnection, may not leave
enough time for  destruction of the field.

The physical model used by all the above authors is, in fact, not 
a reconnection model in the classical sense.
The geometry and dynamics of the flow
in the above models is completely different from
(and in some sense contradictory to)
 the reconnection  models.
In the original Coroniti model, the thickness of the dissipation layer,
which is assumed to be equal to  the Larmor radius based on the external 
magnetic field and internal  thermal particle velocity, 
plays a passive role. As the magnetic field of the  wind decreases
with radius, the thickness of the dissipation layer increases: the 
 reconnection layer ``eats out''  the  magnetic field. At the same time
the inflowing plasma always remains at rest
 with respect to reconnection layer.
This type of a dynamical behavior
 is quite contrary to   reconnection, where magnetic field actually
flows into the reconnection region.
The flow dynamics  proposed by LK
is even more different from the classical reconnection picture. 
LK have argued that the expansion of the
 reconnection layer will  push away  the magnetic field -
a situation  {\it reverse}  to that of reconnection, where magnetic field
is ``sucked'' into  the  reconnection layer.

In this paper we reconsider the problem of reconnection
in  pulsar winds, applying the  Lyutikov \& Uzdensky (2002)
model of relativistic  reconnection.
Lyutikov\& Uzdensky (2002) presented a relativistic generalization of the
simplest model of magnetic reconnection --- the Sweet--Parker
model --- to strongly magnetized plasmas.
In the spirit of the  Sweet--Parker
model, the reconnection layer is assumed to have  a rectangular
shape with  a width $L$ and thickness
$\delta \ll L$.
The width $L$ of the reconnection layer is determined
by the global system size; also prescribed are the ratio $\sigma$
of the  magnetic field energy density to the plasma energy density 
in the ideal-MHD inflow region, and the plasma resistivity $\eta$.
 In contrast, the thickness
$\delta$ of the reconnection region, and the plasma inflow and outflow
 velocity are calculated as a part of the analysis.
As the plasma enters   the reconnection layer,
it slows down, coming to a halt at a stagnation point.
At the same time,  magnetic energy is dissipated and converted into
internal energy of the  pair-rich plasma.
In the out-flowing region, the plasma is accelerated by the
pressure gradients,  reaching some terminal relativistic velocity
$\gamma_{\rm out}$.

Lyutikov\& Uzdensky (2002) have found that the 
structure of the reconnection layer (its thickness, the inflow
and outflow velocities)
depend on the ratio of  two large
dimensionless parameters of the problem - 
magnetization parameter
 $\sigma \gg 1$
and the Lundquist number
\be
S= { L c \over \eta}  \gg 1.
\ee

In the sub-alfvenic regime, $\sigma  \ll S^2$,
the flow is determined by the set of equations
\footnote{These relations are valid for both the non-relativistic
reconnection, $\sigma \ll 1$, and for relativistic, $\sigma \gg 1$.}
\ba &&
 \beta_{\rm in}  \gamma_{\rm in}  \sim { L \over \delta} { 1 \over S},
\nn &&
\gamma_{\rm in}(1+\sigma)   \sim \gamma_{\rm out},
\nn &&
\beta_{\rm in} \sim (1+\sigma) { \delta \over L}.
\ea
The inflow velocity is non-relativistic, $\beta_{\rm in} \ll 1$,
 for $ \sigma \ll S$,  strongly relativistic sub-alfvenic, 
 $ 1 \ll \gamma_{\rm in} \ll \gamma_A $
for $S \ll  \sigma \ll S^2$ ($\gamma_A = \sqrt{2 \sigma}$ is
 the Lorentz
factor of the \Alfven wave  velocity in the incoming region), 
and near alfvenic, $\gamma_{\rm in} \sim  \gamma_A $ for $\sigma \geq S^2$.
The
   outflowing plasma
is moving  always relativistically, $\gamma_{\rm out} \gg 1$ 
if $\sigma \gg 1$.

\section{Reconnection in pulsar winds}

Beyond the light cylinder the pulsar wind   is quasi-radial, moving 
with strongly  relativistic velocity (typical Lorentz factor 
 $\gamma_0 = \sqrt{\sigma} \sim 100$), and is
 strongly magnetized $\sigma \sim 10^4$, dominated by the 
  toroidal magnetic field $B \sim 1/r$.
In the equatorial plane of an oblique rotator
 the  alternating polarity of the magnetic field
create conditions favorable for reconnection
(Michel  1994, Coroniti 1990,  LK).

We  are interested in the maximum possible reconnection rate
in the pulsar wind (conventionally, the term ``reconnection rate''
 refers to the inflow velocity of plasma in terms of \Alfven velocity).
 The maximum reconnection rate
corresponds to the maximum value for the resistivity and thus the
minimal Lundquist number, which
 may be estimated using Bohm's arguments that the
maximum diffusion coefficient
in the magnetized plasma cannot  be much larger than  $r_L v$
where $ r_L  $ is the Larmor radius 
 and $v$ is the typical velocity
of the electrons (of the order  of the speed of light in our case).
Thus, in the limit $\sigma \gg 1 $ 
\be
\eta \sim { c^2 \over \om_B}
\ee
which gives
\ba &&
S \sim { L \over r_L}
\label{SS}
\\ &&
\delta \sim {r_L \over \beta_{\rm in} \gamma_{\rm in}}
\label{SS1}
\\ &&
\beta_{\rm in}^2 \gamma_{\rm in} \sim  \sigma {r_L \over L}
\label{SS2}
\ea

The cyclotron frequency $\om_B$  in the  wind frame
\be 
\om_B \sim { \om_{B, LC} \over \gamma_0} {  r_{LC} \over r} 
\label{ob}
\ee
where $  r_{LC} \sim c/\Omega$ is the pulsar light cylinder,
and $ \om_{B, LC}$ is the  cyclotron frequency at the light
cylinder.

The maximum  width
 of the reconnection layer cannot be larger than  the hydrodynamically 
causally connected sector of the wind $L \sim r/ \gamma_0$.
Using eqns. (\ref{SS}-\ref{ob})  we can then  estimate the Lundquist number 
in the wind
\be
S \sim { \om_{B, LC} \over \gamma_0^2 \Omega}
\label{S}
\ee
Note, that the Lundquist number is independent of radius.

For a ``typical'' pulsar with the surface field $B_{NS}
\sim 10^{12}$ G and $\Omega \sim 10$  rad/sec, $ \om_{B, LC} \sim 10^9$
rad/sec
the Lundquist number is
\be
S\sim 10^4
\ee
and thus for $\sigma \leq 10^4$ we expect a weakly relativistic
inflow velocity with
\be
\beta_{\rm in}^2 \gamma_{\rm in} \sim { \sigma \over S} \sim 1
\label{betain}
\ee

\section{Discussion}

Using the formulation of relativistic reconnection by Lyutikov \& Uzdensky 
(2002)
we have found  that 
the dynamical and kinematic constraints  set on the rates of reconnection
in pulsar winds may still allow  very efficient reconnection 
given by eq. (\ref{betain}). 
If the inflow velocity in the plasma frame  indeed reaches $\sim  c$, 
then  magnetic field of the wind would annihilate  
after  propagating $\sim \gamma_0^2 r_{LC} \sim 10^4 r_{LC}$.
We do not believe that such high rates are  indeed realized
(see, though, Kirk et al. 2002, who have argued that
efficient dissipation within several light cylinders may be 
responsible for pulsar high energy emission). By choosing the
Bohm prescription for resistivity and by neglecting the 
pressure of the ambient plasma  we  have estimated {\it the upper
limits}  on the rates of reconnection.
More realistic calculations should be based on the microphysics of 
the reconnection
layer, e.g. tearing mode in relativistic regimes
(Zelenyi \& Krasnoselskikh 1979) and computer simulations of  relativistic
reconnection layers (Zenitani \& Hoshino 2001, 
Larrabee et al. 2002).  This should 
provide the estimates for the Lundquist number in the layer.
In addition, 
heating of the  plasma by 
dissipating magnetic fields should also be taken into account. 
The 
  corresponding model should be at least two-dimensional, allowing for
the extra pressure to be relieved in the $\theta$ direction.

In spite of these limitations, 
this simple estimate shows that  reconnection can be very  efficient 
in relativistic plasmas. 
In the  case of  the Crab nebula, the magnetic flux has to be destroyed only
by the time the wind reaches 
  the  wisps  located at $r_w\sim 10^{17}$
cm, or $\sim 10^8$  light cylinder radii away. Assuming that initially
$\gamma_0 \sim 100$, the 
inflowing velocity necessary to destroy the magnetic field needs to be
only $10^{-4}$ of the speed of light.
This is 4 orders of magnitude smaller than the maximum reconnection velocity
(\ref{betain}).
To provide such an inflow velocity the Lundquist number may be 8 orders 
of magnitude larger.
Given such  large range of allowed parameters, it is possible  that the 
reconnection indeed is  able to destroy  effectively the magnetic flux
in  pulsar winds. 

Perhaps the main argument against reconnection in the wind comes from the
absence of  observed high energy emission from the central
part of the Crab nebula (Weisskopf et al. 2000). 
Reconnection is usually accompanied by efficient 
particle heating and acceleration  which should result in radiative losses.
In fact, in the case of pulsar winds,  reconnection need not to destroy the 
magnetic field - it need only destroy the magnetic flux
(e.g., by changing the topology of the field lines) to allow the 
flow to match the non-relativistically moving boundary of the nebular.
Changing of topology may require very little dissipation.
Alternatively, the pulsar wisps, conventionally associated with
the reverse shock, may indeed be the signs and the sites
 of reconnection (Michel 1994)
happening in an initially cold  wind  which has not crossed the fast sonic
point (cold winds have asymptotic fast Mach number at most unity,
 Kennel et al. 1983; thus, in  order 
for reconnection to be efficient in slowing down  the pulsar wind
the flow must be  subsonic).

The results of this work  may be contrasted with those of 
Coroniti (1990) and  LK.
On a microscopical level there are some similarities.
For example, 
comparing the thickness of  reconnection sheet 
{\it assumed} by Coroniti and LK
 with our result for the Bohm-type
diffusion,
eq. (\ref{SS1}), shows that the thickness of the reconnection layer does become
of the order of  Larmor radius, $\delta \sim r_L \propto r$,  but
only  when the inflowing velocity becomes 
weakly relativistic  $\beta_{in} \gamma_{in} \sim 1$.

The key difference   is that 
the ``reconnection models'' of Coroniti (1990) and LK
are one-dimensional.  If the flow is forced to be one-dimensional
then the dissipated energy  has  to stay inside the  reconnection layer
 in a thermalized form, leading
to the  unusual dynamics  derived  by LK. 
The assumption of one-dimensionality  is in a sharp contrast to 
classical reconnection  which is at least  a two-dimensional
(and most likely three-dimensional) process. 
Thus, if reconnection happens in pulsar winds the flow
structure will be at least  two-dimensional, so that 
 the particle pressure  created by the  dissipation of the
 magnetic field will  be 
relieved in the direction {\it  orthogonal}  to the initial inflow direction
(radial)
 and not along it as argued by
LK.

\begin{figure}
\psfig{file=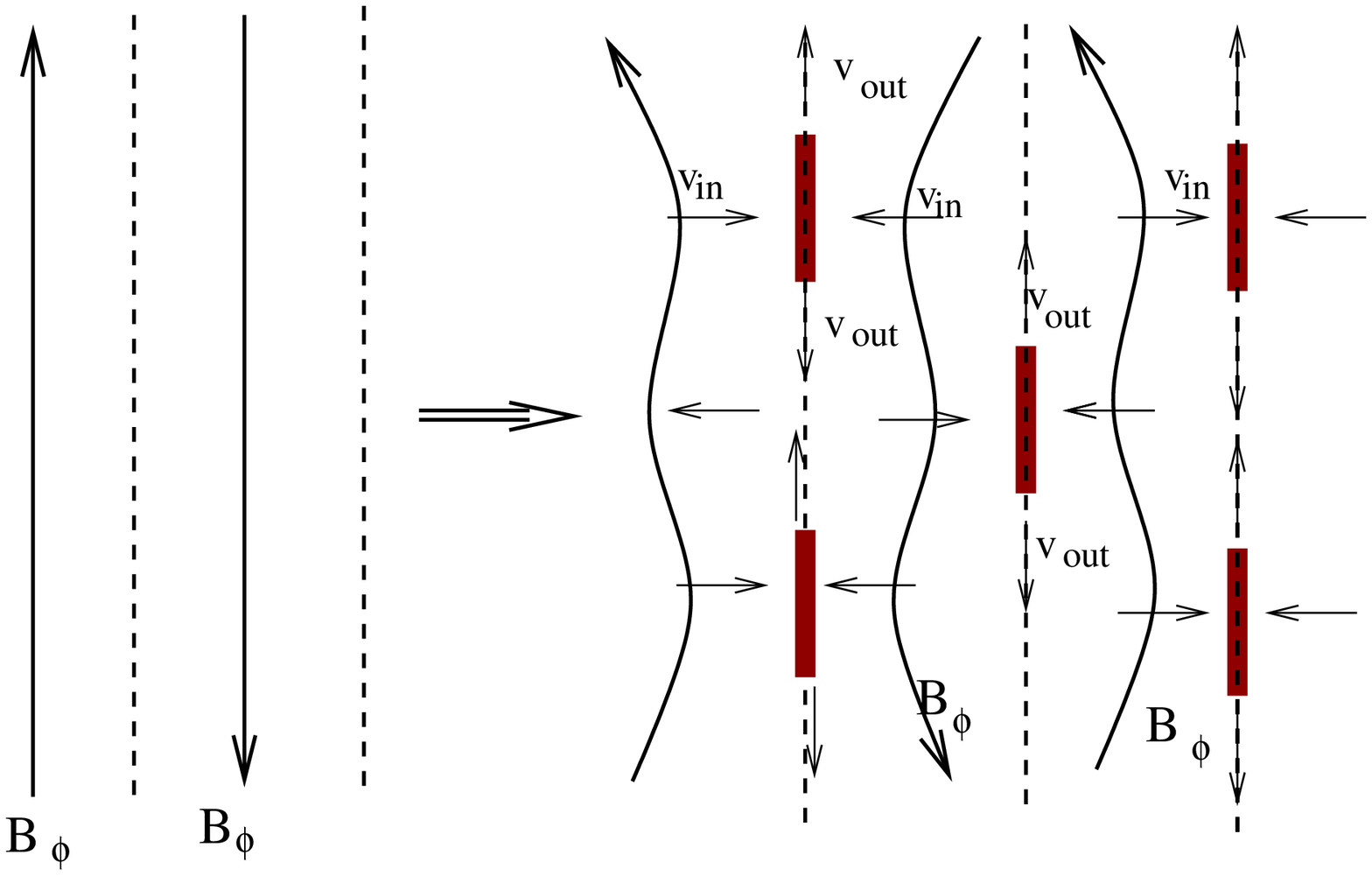,width=17cm}
\caption{
A sketch of  possible two-dimensional reconnection in a  pulsar wind.
Solid lines represent magnetic field, dashed lines - field separatrix, 
 thick boxes - reconnection regions; arrows indicate the direction
of the flow. The
initially azimuthally symmetric striped magnetic field breaks into
separate reconnection regions.
 In the vicinity of 
 each  reconnection region
the structure of the velocity field  resembles a conventional two-dimensional 
Sweet-Parker flow.  
This cartoon serves only to show, that giving up 
azimuthal symmetry, it is possible to have locally 
a classical two-dimensional reconnection picture.
 }
\label{windrec1}
\end{figure}

In the  two-dimensional Sweet-Parker model the outflowing velocity is directed
along the external magnetic field lines. In  the case of the
 equatorial flow in 
 pulsar winds, this 
corresponds to the azimuthal  direction. In a perfectly 
axisymmetric picture such outflow is obviously  impossible. 
Contrary to LK  we suggest that instead of preserving 
 the axially symmetric form at all costs, the flow would become
non-axially symmetric and thus two-dimensional (and most likely
three dimensional). 
As soon as 
reconnection starts in some localized region,
a localized deposition of energy will distort the flow and break
the azimuthal symmetry, setting up some complicated velocity pattern,
with plasma moving in and out of  the
 reconnection regions (see Fig. \ref{windrec1}).
Mutual interaction of different layers then becomes an important issue.
 Redeposition of  energy back into the flow from reconnection regions
would   affect the rates of reconnection only after a considerable fraction
of the magnetic field has been dissipated.  Under certain circumstances
reconnection in one localized region may  push plasma 
into other  reconnection sites, speeding up  reconnection,
as is illustrated in Fig. \ref{windrec1}.
 These issues are 
 beyond the scope of this Letter.

Based on these arguments  we conclude that the 
 one-dimension approach to the problem of reconnection in 
 pulsar winds  and the ensuing result that reconnection is not important
is likely to be incorrect - one has to consider
at least two-dimensional (and possible full three-dimensional)
problem.
Our simple estimates show that 
 in pulsar winds the plasma  may flow
into the reconnection region with  mildly relativistic
velocities insuring a much more efficient 
reconnection than argued by LK.
Since we have calculated only the upper limits on the reconnection rate,
we cannot make a conclusive statement if reconnection indeed
occurs efficiently. This requires an understanding
 of the microphysical
processes  (e.g., evolution of tearing mode) in strongly relativistic 
magnetized plasmas and of the three-dimensional structure of the
flow.

\begin{acknowledgements}
We  would like to thank John Kirk, Yuri Lyubarsky
 for their interest in this work and Vicky Kaspi and 
Alissa Nedossekina for comments on the manuscript.
\end{acknowledgements}

\end{document}